# Magneto-optical spectra of an organic antiferromagnet as a candidate for an altermagnet

Satoshi Iguchi,[1,*] Hiroki Kobayashi,[1] Yuka Ikemoto,[2] Tetsuya Furukawa,[1] Hirotake Itoh,[3] Shinichiro Iwai,[3] Taro Moriwaki,[2] and Takahiko Sasaki[1]

[1]*Institute for Materials Research, Tohoku Univ., Aoba-ku, Sendai 980-8577, Japan*
[2]*Japan Synchrotron Radiation Research Institute, SPring-8, Sayo, Hyogo 679-5198, Japan*
[3]*Department of Physics, Tohoku Univ., Aoba-ku, Sendai 980-8578, Japan*

We have measured the magneto-optical Kerr effect (MOKE) in an orthorhombic organic antiferromagnet κ-(BEDT-TTF)$_2$Cu[N(CN)$_2$]Cl (κ-Cl), which is a candidate for an altermagnet. From the Maxwell equations, we derived matrix-type general formulae describing the optical propagation and reflection for arbitrary crystals. These formulae enabled us to correctly measure and obtain the off-diagonal optical responses of κ-Cl. The MOKE of κ-Cl appeared at around the Néel temperature and exhibited a nonlinear field dependence in the antiferromagnetic phase. This nonlinear field dependence eliminates a simple origin due to the net canted magnetization. The obtained off-diagonal optical conductivity spectra for the entire π-electron band clearly show three features. One is the large peaks at the spectral ends due to the magnetic origin with the large energy scale compared to the very small spin-orbit interaction in κ-Cl. The others are the middle region spectra proportional to the diagonal conductivities, possibly related to the symmetric piezomagnetic effect and the standard antisymmetric origins. These results suggest the altermagnetic response of κ-Cl. We also discuss the analogy between the magnetization in ferromagnets and the net magnetization and Néel vector with respect to the magneto-optical configurations.

## I. INTRODUCTION

The anomalous Hall effect (AHE) is originally a phenomenon in which the electric current curves when macroscopic magnetization occurs. The typical origin has been quantum theoretically understood by the Berry curvatures or magnetic monopoles (anti-crossing bands) in the momentum-space caused by the spin-orbit coupling (SOC) [1,2]. Such generalizations and simple understandings generate new physical insights. The spin-orbit interaction and macroscopic magnetization are not necessary for the spontaneous current bending. Magnetic monopoles in *k*-space can emerge not only due to SOC, but also due to scalar spin chirality [3]. In addition, macroscopic magnetization is not required if the time-reversal symmetry is broken. For example, frustrated noncoplanar spins can cause a non-zero net scalar spin chirality without net magnetization [4]. Furthermore, coplanar spins can also break the time-reversal symmetry [5–8].

In recent years, several groups [9,10] have proposed that AHE or magneto-optical Kerr effect (MOKE) can appear even under a complete antiparallel spin order without net magnetization. Such collinear antiferromagnetic (AF) materials have $sk_xk_y$ -type spin-splitting bands and classified as altermagnets [11,12], where $s$ and $k$ are spin and wavenumber, respectively. Altermagnet is different from the standard antiferromagnet with the identical up and down spin bands. Therefore, the term $sk_xk_y$ is considered as the order parameter of the altermagnet. The altermagnets are expected to show interesting phenomena and properties such as new efficient spin current sources [13,14] and piezomagnetic effects [15,16] and have therefore attracted attention in recent years [11,17]. The altermagnetic spin band splitting, as well as the spin current generation and piezomagnetic effect, can appear without relativistic spin-orbit interactions, which is advantageous for applications. However, except for the topological origin with non-coplanar spins [18,19], the SOC is necessary for the emergence of AHE/MOKE [20]. $sk_xk_y$ -type spin bands have been observed in some materials such as MnTe [21] and RuO$_2$ [22] by XMCD, which detects the same symmetry as AHE/MOKE, and both theory and experiment reveal the spin-band splitting characteristic of the altermagnetism. On the other hand, MOKE is able to determine the off-diagonal optical conductivity usually in infrared and visible light regions, which is directly related to the transport properties. Therefore, MOKE is expected to reveal magneto-transport properties of altermagnets in a complementary way to XMCD. In this study, we have experimentally revealed the magneto-optical spectra of κ-(BEDT-TTF)$_2$Cu[N(CN)$_2$]Cl (κ-Cl), a candidate organic altermagnet for which these effect have been discussed from the earliest stage [9,13].

κ-Cl has a quasi-two-dimensional layered structure in which the layer of BEDT-TTF molecules responsible for conduction and the monovalent anion layers are stacked alternately as shown in Fig. 1(a) [23]. The BEDT-TTF molecules dimerize, and hole carriers existing in the dimer orbital are strongly correlated, resulting in the phase transition from paramagnetic metal to antiferromagnetic dimer-Mott insulator at about 25 K [24–26] (as shown in Figs.1(b) and 1(c)). In this AF phase, the Dzyaloshinsky-Moriya (DM) interaction works between the spins since there is no inversion center between the two dimers, resulting in the spin-canted AF state [26] [Fig.1(b)]. This spin canting component causes a small macroscopic magnetization (0.003 $\mu_B$/dimer), as clearly seen in Fig.1(c).

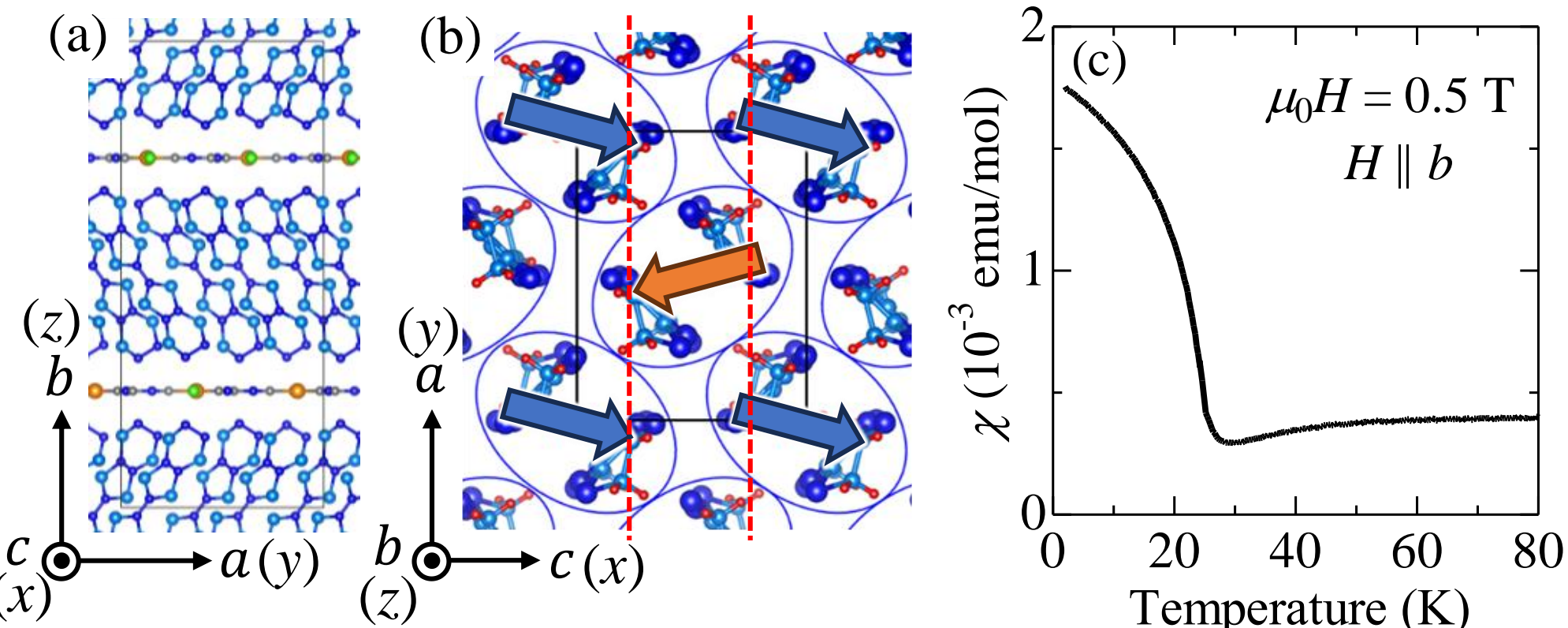

FIG 1. (a) Crystal structure of the inter-layer structure viewed from the *c*-axis. (b) Conduction layer in $\kappa$-Cl. Ovals and arrows represent the dimerized BEDT-TTF molecules and spins at the dimer in the canted antiferromagnetic phase, respectively. The broken lines represent glide mirror planes perpendicular to the *c*(*x*)-axis, which is lost by AF spin ordering. (c) Temperature dependence of static magnetic susceptibility in $H||b$ (perpendicular to the conduction layer) at 0.5 T. The sudden increase of the susceptibility at 25 K shows the canted antiferromagnetic transition. The *x*-, *y*, and *z*-axes corresponding to the *c*-, *a*-, and *b*-axes are shown in (a) and (b), respectively.

To explain why κ-Cl is a candidate altermagnet in this AF state, we will use Table 1. Tables 1 show the character tables of (a) the point group $C_{2v}$ as a minimum model [27,28] and (b) the magnetic point group $C'_{2v}$ $(m'm2')$, which correspond to the paramagnetic and an AF phases with canted spins along the *y*(*a*)-axis in κ-Cl, respectively. Due to the AF spin order, the glide symmetry shown by the broken lines in Fig. 1(b) was lost, which corresponds to the change in the symmetry operation from $m_x$ in $C_{2v}$ to $Tm_x$ in $C'_{2v}$ [29]. The *c*(*x*)-axis is the easy axis of the AF spins drawn as blue and orange arrows in Fig. 1(b). These antiparallel spins, i.e., the two spin sublattices, are interchanged by a half of the symmetry operations, $Tm_x$ and $m_y$**.** This spin-sublattice segmentation corresponds to the irreproducible representation (irrep) $A_2$ in $C'_{2v}$, where the characters of $Tm_x$ and $m_y$ are -1 and the basis $M_{x(c)}$ belongs to. The symmetry of $M_{x(c)}$ is the same as the (*x*-component of) Néel vector in this system. Since $m_y$ can be devided into the multiplication $(Tm_x)(TC_{2z})$, the subset $\{Tm_x, m_y\} = (Tm_x)H$, where $H = \{E,\ TC_{2z}\}$ is a normal subgroup and interchanges the spins *within* the sublattice. Therefore, $C'_{2v} = H + (Tm_x)H$.

The wavenumber terms $k_x$, $k_y$, and $k_z$, as well as momentum and velocity, i.e., the time derivative of *x*, do not usually appear as bases. However, we have added these terms for the sake of a rough explanation to grasp the nature of altermagnetism and its relationship to related phenomena.

In the AF phase represented by Table 1(b), the products of bases $\{M_x,\ xy,\ k_xk_y\}$ in the one-dimensional irrep $A_2$ such as $M_xk_xk_y$ and $M_xxy$ ($A_2 \times A_2$), belong to the totally symmetric $A_1$ irrep, in addition to $M_y$. The same holds for the $B_1$ and $B_2$ irreps. However, since $M_x$ is the largest spin component in the AF phase of κ-Cl, the basis products including $M_x$ belonging to $A_2 \times A_2 (= A_1)$ are larger than that for $B_2 \times B_2 (= A_1)$ including $M_z$. Therefore, $M_xk_xk_y$ and $M_xxy$ are the dominant order parameters, i.e. magnetic octupoles [30,31], of altermagnetism and piezomagnetic effects, respectively. Since $k$ has the same symmetry as electric current $j$, the order parameter $M_xk_xk_y$ can be reconstructed as $(M_xk_x)k_y \sim (s_xj_x)j_y = (s_xj_y)j_x$, where $s_xj_{x(y)}$ means spin current generated by the electric current

TABLE 1. Character table and basis of (a) point group $C_{2v}$ and (b) magnetic point group ($C'_{2v}$ or $m'm2'$) with spin canting along the *y*-axis as represented by $M_y$ in the irreducible representation $A_1$. Symmetry operations $E, C_{2z}, m$ and $T$ represent identity, rotation, mirror, and time-reversal operation, respectively. $C_{2v}$ is a normal subgroup of $D_{2h} (= C_{2v} \times I)$ which corresponds to the overall crystal structure of κ-Cl, where $I$ represents the space inversion. Because of omitting $I$, bases changing its sign with respect to $I$, such as *x*, *y*, and *z*, are not shown. The coordinates $(x,\ y,\ z)$ correspond with the crystal axes $(c,\ a,\ b)$ as in Fig. 1, respectively. $(k_x,\ k_y,\ k_z)$ is considered a function with the same symmetry as momentum, velocity, or electric current, i.e., the time derivative of $(x, y,\ z)$, respectively. However, the product of $k$s, e.g., $k_xk_y$ is time-reversal even and belongs to the same irreproducible representation as *xy*. Since (a) $C_{2v}$ is irrelevant to time-reversal operations, bases $M$ are represented with parentheses.

(a)

| $C_{2v}$ | $E$ | $C_{2z}$ | $m_y$ | $m_x$ | basis | | |
|---|---|---|---|---|---|---|---|
| $A_1$ | 1 | 1 | 1 | 1 | | | |
| $A_2$ | 1 | 1 | -1 | -1 | $(M_z)$ | $xy$ | $k_xk_y$ |
| $B_1$ | 1 | -1 | 1 | -1 | $(M_y)$ | $zx$ | $k_zk_x$ |
| $B_2$ | 1 | -1 | -1 | 1 | $(M_x)$ | $yz$ | $k_yk_z$ |

(b)

| $C'_{2v}$ | $E$ | $TC_{2z}$ | $m_y$ | $Tm_x$ | basis | | |
|---|---|---|---|---|---|---|---|
| $A_1$ | 1 | 1 | 1 | 1 | $M_y$ | | |
| $A_2$ | 1 | 1 | -1 | -1 | $M_x$ | $xy$ | $k_xk_y$ |
| $B_1$ | 1 | -1 | 1 | -1 | | $zx$ | $k_zk_x$ |
| $B_2$ | 1 | -1 | -1 | 1 | $M_z$ | $yz$ | $k_yk_z$ |

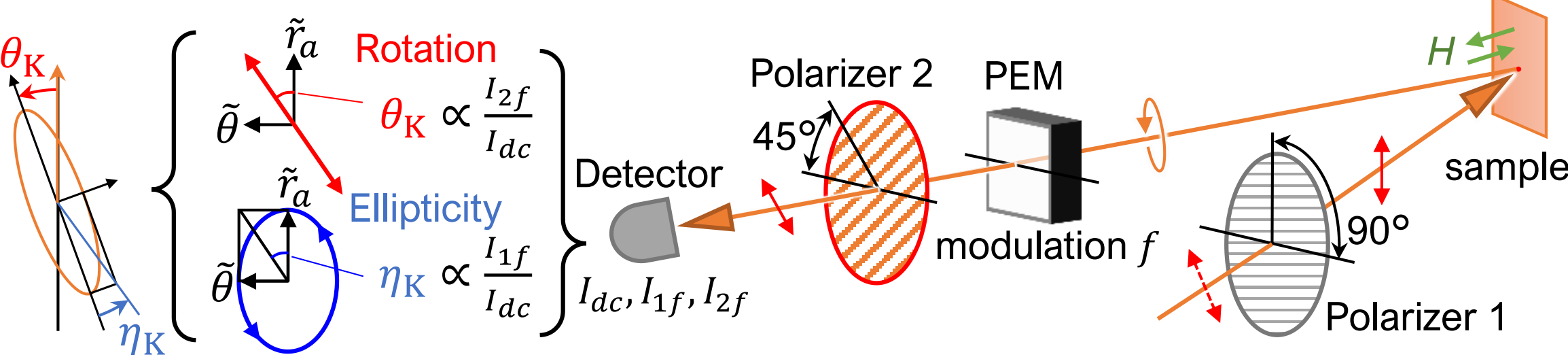

FIG 2. Schematic diagram of the optical system and signal analysis. Polar Kerr configuration of near normal condition. The crystal surface is the *ca* plane [see Fig. 1(b)]. The incident angle averaged is estimated to be about 9°. The reflected light is modulated by PEM at frequency $f$. Electrical signals are decomposed into the dc, $1f$, and $2f$ components by a lock-in-amplifier, and each spectrum is obtained by FTIR. The two angles, Kerr rotation $\theta_{\mathrm{K}}$ and ellipticity $\eta_{\mathrm{K}}$, correspond to the real (in-phase) and imaginary (90° delay) part of the ratio $\tilde{\theta}/\tilde{r}_a$.

$j_{y(x)}$, respectively. These spin currents will show "symmetric" directional dependence in contrast to the antisymmetric spin Hall effect, as theoretically suggested in Ref. [13].

In this canted-AF state, Naka et al. [9] predicted that κ-Br [32], which has the same intralayer structure as κ-Cl but has a different stacking structure, can show the AHE due to the altermagnetism at zero magnetic field ($H$). In κ-Cl, the effect cancels out between the two layers at zero field, but half of the spins are flipped (rotated by 180°) during the spin-flop at about 0.4 T in $H||b$ [25,26], so all the spins reverse between $\mu_0 H > 0.4$ T and $\mu_0 H < -0.4$ T, and AHE can occur by this mechanism. Note that the net magnetization [i.e., $M_y$ for Table 1(b)] is irrelevant to the mechanism of the altermagnetism since the primary order parameter is $M_x k_x k_y$. Rather, the SOC is essential for the AHE/MOKE emergence.

As mentioned above, in $\kappa$-Cl, the AF spin order breaks the glide symmetry, which breaks the left-right symmetry of the current. In general, the SOC is necessary for the generation of AHE/MOKE even in altermagnets [20]. In κ-Cl/Br, requires the Zeeman type SOC in addition to the AF spin ordering [9]. This AHE/MOKE can manifest even in the absence of the spin canting of $s_{a(y)}$ and $s_{b(z)}$. While the spin canting is unnecessary, the symmetry of the AHE/MOKE appearance is equivalent to that of the system with a spontaneous magnetization along the $b(z)$-direction in $H \parallel b$ [9,20]. Therefore, this is a very curious phenomenon, although equivalent results have been reported for perovskite systems with the same *Pbnm* (or *Pnma*) space group [31,33]. In $\kappa$-Cl, the Zeeman type SOC along the $c(x)$-direction interacts with the spin component $s_c$, resulting in a change in the electron's energy, site density, and transfer to the adjacent sites depending on the $\pm c$-spin direction. The group theoretical multipole considerations [27,34] suggest that κ-Cl can exhibit AHE because the pattern of the antiparallel spin order and the spin-orbit interaction at each site mesh well in a coordinated manner as well as the anisotropic electron transfers. This symmetry is a distinctive feature of the mechanism.

Of course, AHE is a suitable phenomenon to explore the functionality in altermagnetism, but it cannot be measured in insulators. However, AF metals are rare even in altermagnet candidates [10,11,35], and κ-Cl is no exception [36,37]. In such AF insulators, the magneto-optical Faraday and Kerr effects, which are AHE in the optical frequency domain, are appropriate detectors for revealing the origin of AHE [38–43], which may include the energetic information of the altermagnetic spin-band separation, although the other contributions may be detected [42,44]. We measured the MOKE spectrum of κ-Cl in the infrared light region, in which κ-Cl shows major optical transitions related to π electrons [45,46].

## II. METHODS

### A. Outline

The quantities obtained by MOKE measurements are two scalar angles, Kerr rotation $\theta_{\mathrm{K}}$ and ellipticity $\eta_{\mathrm{K}}$. As shown in Fig. 2, when an incident linearly polarized light is reflected by a magnetic material, the polarization of reflected light rotates and elliptically expands. These angles can be measured using polarizers and/or a photoelastic modulator (PEM). In an isotropic crystal the off-diagonal dielectric constant $\tilde{\varepsilon}_{xy}$ can be obtained from the measured quantities $(\theta_{\mathrm{K}}, \eta_{\mathrm{K}})$ by the following equations, assuming that the scalar complex reflectance $\tilde{r}$ or refractive index $\tilde{n}$ is known.

$$\tilde{\theta}_{\mathrm{K}} \equiv \theta_{\mathrm{K}} + i\eta_{\mathrm{K}} = \frac{\tilde{\varepsilon}_{xy}}{\sqrt{\tilde{\varepsilon}_{xx}}(1-\tilde{\varepsilon}_{xx})},$$

$$\tilde{\varepsilon}_{xx} = \tilde{n}^2, \qquad \tilde{n} = \frac{1-\tilde{r}}{1+\tilde{r}}. \qquad (1)$$

Hereafter, tilde represents complex number, and dielectric constant $\tilde{\varepsilon}$ and magnetic permeability $\tilde{\mu}$ (used later) are relative ones.

Since κ-Cl is orthorhombic (space group *Pbnm*) [23], unfortunately the above equation cannot be used. Anisotropic materials such as κ-Cl/Br exhibit birefringence and linear dichroism. These optical effects due to longitudinal (diagonal) responses are about 100-1000 times larger than transverse (off-diagonal) responses. This significant difference is similar to the ratio between the longitudinal resistivity $\rho_{xx}$ and the (anomalous) Hall resistivity $\rho_{yx}$. Therefore, although the orthorhombic crystals have higher symmetry than monoclinic or triclinic ones, we need to pay careful attention to the separation of the diagonal and off-diagonal effects. For this purpose, we have to correctly measure the MOKE and correctly analyze the off-diagonal dielectric constant $\tilde{\varepsilon}_{xy}$.

This difficulty can also be seen in studies on the Faraday effect of orthoferrites in the 1960s and 70s [47–49]. From these studies, we can learn that the Jones's method [50,51] is easier to understand and calculate, where the electric polarization of light $\vec{E}$ is expressed as a two-dimensional vector and the transmittance, reflectance, and optical elements are expressed as 2×2 matrices.

We consider the *xy* plane of an orthorhombic magnet in the Faraday configuration for (near) normal incident, where magnetic field is perpendicular to the sample surface (see Fig. 2). The two diagonal elements $\tilde{r}_\alpha = r_\alpha + is_\alpha$ $(\alpha = x, y)$ in the 2×2 complex reflectance matrix $r$ are different in the orthorhombic crystal. The two-remaining off-diagonal elements $\pm\tilde{\theta} = \pm(\theta + i\eta)$ corresponding to magnetic effects have opposite signs, which can be understood shortly after, considering the Onsager's reciprocal theorem,

$$r = \begin{pmatrix} \tilde{r}_x & -\tilde{\theta} \\ \tilde{\theta} & \tilde{r}_y \end{pmatrix}. \tag{2}$$

When linearly polarized light of $E||y$ is incident on this crystal as shown in Fig. 2, the reflected light becomes elliptically polarized in general, and the complex angle $\tilde{\theta}_{\mathrm{K}|y} = \tilde{\theta}/\tilde{r}_y$ is expressed as the ratio of the electric field of reflected/incident light,

$$r\begin{pmatrix} 0 \\ 1 \end{pmatrix} = \begin{pmatrix} -\tilde{\theta} \\ \tilde{r}_y \end{pmatrix} = \tilde{r}_y \begin{pmatrix} -\tilde{\theta}_{\mathrm{K}|y} \\ 1 \end{pmatrix}. \tag{3}$$

From this equation, we notice that when compared with the $E||x$ incidence, the Kerr rotations and ellipticities depend on the polarization directions of the incident light ($\tilde{\theta}_{\mathrm{K}|x} \neq \tilde{\theta}_{\mathrm{K}|y}$). The off-diagonal reflectivity $\tilde{\theta}$, a material-specific quantity, is obtained only by calculating $\tilde{\theta} = \tilde{\theta}_{\mathrm{K}|y}\tilde{r}_y = \tilde{\theta}_{\mathrm{K}|x}\tilde{r}_x$.

Even if the off-diagonal element $\theta$ in the $r$ matrix is experimentally obtained, there has not been a simple method to obtain the off-diagonal optical permittivity $\tilde{\varepsilon}_{xy}$ from it. In recent years, we have found the calculation formula [52], which is, however, almost the same as Eq. (1) only with the use of $2 \times 2$ matrices $I, r, n,$ and $\varepsilon$.

$$\begin{aligned} \tilde{\varepsilon}_{xy} = -\tilde{\varepsilon}_{yx} &= \frac{\tilde{\theta}(1+\tilde{n}_x)(\tilde{n}_x+\tilde{n}_y)(\tilde{n}_y+1)}{2}, \\ \varepsilon = n^2, &\qquad n = (I-r)(I+r)^{-1}, \end{aligned} \tag{4}$$

where $I$ represents the identity matrix. This similarity is also convincing from the physical requirement that Eqs. (4) should be reduced to Eqs. (1) for isotropic materials. Equations (4) are general formulae that can be applied to arbitrary samples under the normal incidence condition.

It is important to note that Eqs. (4) are not the result of substituting matrices to the corresponding scalar quantities in Eqs. (1), though it is mathematically impossible. In the next subsection, we will present a rigorous derivation of the formulae in Eqs. (4), which was not explicitly demonstrated in the previous paper [52].

### B. Formula derivation

Assuming crystals with lower symmetries, consider the $3 \times 3$ general (relative) dielectric constant $\varepsilon_3$ and magnetic permeability $\mu_3$ tensor as,

$$\varepsilon_3 = (\tilde{\varepsilon}_{ij}), \qquad \mu_3 = (\tilde{\mu}_{ij}), \qquad (i, j = \alpha, \beta, \gamma). \tag{5}$$

Hereafter, the diagonal elements are denoted as $\tilde{\varepsilon}_{ii} = \tilde{\varepsilon}_i, \tilde{\mu}_{ij} = \tilde{\mu}_i$. The key idea in proceeding with the derivation of Eq. (4) is to rewrite the Maxwell equations into $2 \times 2$ matrix equations with the measurable $2 \times 2$ reflectance matrix in mind.

Assuming that the surface of the crystal is the $\alpha\beta$ plane and the $\alpha$- and $\beta$-axes are parallel to the *x*- and *y*-directions in the coordinate system, respectively $(\alpha||x, \beta||y)$. The light traveling in the $z$-direction with a wavenumber $k_z$ is generally written as, $\boldsymbol{E} = \boldsymbol{E}_0 \exp i(k_z z - \omega t)$. To derive the wave equation for this condition, use the Maxwell equations,

$$\begin{gathered} k_z \cdot D_z = 0, \qquad k_z \cdot B_z = 0, \\ \boldsymbol{k} \times \boldsymbol{E} = \omega\boldsymbol{B}, \qquad \boldsymbol{k} \times \boldsymbol{H} = -\omega\boldsymbol{D}. \end{gathered} \tag{6}$$

$k_z \cdot D_z = 0$ (thus $D_z = 0$) and relations between the field and flux density,

$$\boldsymbol{D} = \varepsilon_0\varepsilon_3\boldsymbol{E}, \qquad \boldsymbol{B} = \mu_0\mu_3\boldsymbol{H}, \tag{7}$$

yield $E_z = -\varepsilon_\gamma^{-1}(\varepsilon_{\gamma\alpha}E_x + \varepsilon_{\gamma\beta}E_y)$. Substituting this $E_z$ into Eq. (7), we obtain a two-dimensional equation for $E_x$ and $E_y$ as,

$$\boldsymbol{D} = \varepsilon_0 \begin{pmatrix} (\varepsilon_\alpha - \varepsilon_{\alpha\gamma}\varepsilon_\gamma^{-1}\varepsilon_{\gamma\alpha})E_x + (\varepsilon_{\alpha\beta} - \varepsilon_{\alpha\gamma}\varepsilon_\gamma^{-1}\varepsilon_{\gamma\beta})E_y \\ (\varepsilon_{\beta\alpha} - \varepsilon_{\beta\gamma}\varepsilon_\gamma^{-1}\varepsilon_{\gamma\alpha})E_x + (\varepsilon_\beta - \varepsilon_{\beta\gamma}\varepsilon_\gamma^{-1}\varepsilon_{\gamma\beta})E_y \\ 0 \end{pmatrix},$$

therefore,

$$\begin{aligned} \vec{D} &= \begin{pmatrix} D_x \\ D_y \end{pmatrix} \\ &= \varepsilon_0 \begin{pmatrix} \varepsilon_\alpha - \varepsilon_{\alpha\gamma}\varepsilon_\gamma^{-1}\varepsilon_{\gamma\alpha} & \varepsilon_{\alpha\beta} - \varepsilon_{\alpha\gamma}\varepsilon_\gamma^{-1}\varepsilon_{\gamma\beta} \\ \varepsilon_{\beta\alpha} - \varepsilon_{\beta\gamma}\varepsilon_\gamma^{-1}\varepsilon_{\gamma\alpha} & \varepsilon_\beta - \varepsilon_{\beta\gamma}\varepsilon_\gamma^{-1}\varepsilon_{\gamma\beta} \end{pmatrix} \begin{pmatrix} E_x \\ E_y \end{pmatrix} \\ &= \varepsilon_0\varepsilon\vec{E}. \end{aligned} \tag{8}$$

Hereafter we will use the arrow instead of bold to represent the two-dimensional vectors. The magnetic permeability can also be reduced to a $2 \times 2$ matrix in the same way.

$$\begin{aligned} \vec{B} &= \begin{pmatrix} B_x \\ B_y \end{pmatrix} \\ &= \mu_0 \begin{pmatrix} \mu_\alpha - \mu_{\alpha\gamma}\mu_\gamma^{-1}\mu_{\gamma\alpha} & \mu_{\alpha\beta} - \mu_{\alpha\gamma}\mu_\gamma^{-1}\mu_{\gamma\beta} \\ \mu_{\beta\alpha} - \mu_{\beta\gamma}\mu_\gamma^{-1}\mu_{\gamma\alpha} & \mu_\beta - \mu_{\beta\gamma}\mu_\gamma^{-1}\mu_{\gamma\beta} \end{pmatrix} \begin{pmatrix} H_x \\ H_y \end{pmatrix} \\ &= \mu_0\mu\vec{H}. \end{aligned} \tag{9}$$

Similarly, the cross product by $\boldsymbol{k}$ becomes a $2 \times 2$ matrix because $k_x = k_y = 0$.

$$(\boldsymbol{k}\times)=\begin{pmatrix}0&-1&0\\1&0&0\\0&0&0\end{pmatrix}k_z$$

$$\rightarrow(\vec{k}\times)=\begin{pmatrix}0&-1\\1&0\end{pmatrix}\hat{k}_z\equiv R\hat{k}_z, \tag{10}$$

where $R=-R^{-1}$ represents a two-dimensional 90 degrees rotation matrix. As well as $R$, $\mu$ and $\varepsilon$ are assumed to be noncommutative and regular matrices. Note that $\hat{k}_z(=-i\partial/\partial z)$ is a scalar operator which acts on $\vec{E}=\vec{E}\exp(ik_z z)$ (at $z=0$).

Using the three $2\times 2$ matrices $(\varepsilon,\mu,R)$, the latter two of Eq. (6) change into,

$$R\hat{k}_z\vec{E}=\omega\mu_0\mu\vec{H},\qquad R\hat{k}_z\vec{H}=-\omega\varepsilon_0\varepsilon\vec{E}. \tag{11}$$

Eliminating the magnetic field by substituting $\vec{H}=(\omega\mu_0\mu)^{-1}R\hat{k}_z\vec{E}$ from the first into the second equation yields a two-dimensional electric field equation,

$$\hat{n}_z^2\vec{E}=(R\mu R^{-1})\varepsilon\vec{E}\equiv\mu_R\varepsilon\vec{E}, \tag{12}$$

where two scalar relationships $\mu_0\varepsilon_0=c^{-2}$ and $\hat{k}_z=\hat{n}_z\omega/c$ are used, and finally $R\mu R^{-1}$ are defined as $\mu_R$, representing the permeability matrix parallel to $\vec{H}$ or rotated by 90 degrees with respect to $\vec{E}$. Equation (12) in the form of 2×2 eigenequation is a general form of the wave equation for light traveling in the $z$-direction in a crystal. When Eq. (12) is diagonalized, the eigenvalues are the square of refractive index, $n_1^2$ and $n_2^2$, and eigenvectors are the corresponding electric fields, namely, light polarizations. In the wave equation, we can define the matrix of refractive index $n$ as,

$$\mu_R\varepsilon=\begin{pmatrix}\mu_\beta\varepsilon_\alpha & \mu_\beta\varepsilon_{\alpha\beta}-\mu_{\beta\alpha}\varepsilon_\beta\\ \mu_\alpha\varepsilon_{\beta\alpha}-\mu_{\alpha\beta}\varepsilon_\alpha & \mu_\alpha\varepsilon_\beta\end{pmatrix}\equiv n^2, \tag{13}$$

where we have made an approximation up to the first order of off-diagonal elements, which can be applied to orthorhombic crystals.

Since the eigen electric field (polarization) $\vec{E}_i$ $(i=1,2)$ corresponding to each of the two refractive indices $n_i$ is obtained, arbitrary electric field of light $\vec{E}_\mathrm{t}$ in the crystal is expressed as a superposition of eigen polarizations $\vec{E}_i$ with $\alpha_i$ as the scalar coefficients,

$$\vec{E}_\mathrm{t}=\vec{E}_1\alpha_1+\vec{E}_2\alpha_2=(\vec{E}_1,\vec{E}_2)\begin{pmatrix}\alpha_1\\ \alpha_2\end{pmatrix}\equiv E\vec{\alpha}, \tag{14}$$

where $E$ is the $2\times 2$ matrix for diagonalizing $n^2$ and $n$ matrix. Writing scalar constants on the right side of the vector makes it easier to deform such equations, as well as to distinguish constants from operators. As mentioned above, $\hat{k}_z$ $(\propto\hat{n}_z\propto\partial/\partial z)$ does not act on scalar or matrix constants such as $r$, but only on the electric field (vector), where the eigenvalue $n_i$ (scalar) is determined for each $\vec{E}_i$ as $\hat{n}_z\vec{E}_i=\vec{E}_i n_i$.

$$\hat{n}_z\vec{E}_t=\vec{E}_1 n_1\alpha_1+\vec{E}_2 n_2\alpha_2$$

$$=(\vec{E}_1,\vec{E}_2)\begin{pmatrix}n_1&0\\0&n_2\end{pmatrix}\begin{pmatrix}\alpha_1\\ \alpha_2\end{pmatrix}\equiv EN\vec{\alpha}, \tag{15}$$

where $N$ is the refractive index matrix diagonalized by the two eigen polarizations. $\vec{\alpha}$ is a vector of coefficients arranged vertically and has no physical meaning. The general matrix form of the refractive index can be obtained by reversing the diagonalization as,

$$n=ENE^{-1}\ \ (n^2=EN^2E^{-1}). \tag{16}$$

This $n$ is the matrix representation of the operator $\hat{n}_z$. Due to the physical constraint of the refractive index ($\mathrm{Re}\{n_i\}\geq 1,\mathrm{Im}\{n_i\}>0$), Eq. 16 is obtained. Using these eigen polarizations and values, we can concisely derive the general formula for reflectance.

The boundary condition at the crystal surface $(x,y)$ at $z=0$ is the continuity of the surface-parallel $\vec{E}$ and $\vec{H}$ on the front and back side. Assuming that the incident light $\vec{E}_\mathrm{in}$ and the reflected light $\vec{E}_\mathrm{r}=r\vec{E}_\mathrm{in}$ are in a vacuum, thus $k_0c=\omega n_0,\varepsilon=1,\mu=1,n_0=1$, and the electric field in the crystal is $\vec{E}_\mathrm{t}=E\vec{\alpha}$ from Eq. (14), then the continuity condition for the electric field $\vec{E}_\mathrm{in}+\vec{E}_\mathrm{r}=\vec{E}_\mathrm{t}$ is,

$$(I+r)\vec{E}_\mathrm{in}=E\vec{\alpha}. \tag{17}$$

Similarly, since the magnetic field is represented as a function of electric field $\vec{H}=\mu^{-1}Rn_z\vec{E}\mu_0^{-1}c^{-1}$ from the first equation of Eqs. (11), $\vec{H}_\mathrm{vac}=\vec{H}_\mathrm{in}+\vec{H}_\mathrm{r}$ on the vacuum side is,

$$\vec{H}_\mathrm{vac}=R(I-r)\vec{E}_\mathrm{in}\mu_0^{-1}c^{-1}, \tag{18}$$

note that the wavenumber of the reflected light is reversed $-\hat{k}_z(\propto-\hat{n}_z)$. As for the magnetic field of transmitted light $\vec{H}_\mathrm{t}$ in the crystal, using Eq. (15),

$$\vec{H}_t=\mu^{-1}R(\hat{n}_z\vec{E}_\mathrm{t})\mu_0^{-1}c^{-1}=\mu^{-1}R(EN\vec{\alpha})\mu_0^{-1}c^{-1}. \tag{19}$$

The continuity condition is $\vec{H}_\mathrm{vac}=\vec{H}_\mathrm{t}$, which becomes,

$$R(I-r)\vec{E}_\mathrm{in}=\mu^{-1}REN\vec{\alpha}. \tag{20}$$

Substituting $\vec{\alpha}$ in Eq. (17) into the above Eq. (20) yields,

$$R(I-r)\vec{E}_\mathrm{in}=\mu^{-1}RENE^{-1}(I+r)\vec{E}_\mathrm{in}. \tag{21}$$

Since this is satisfied for an arbitrary incident light $\vec{E}_\mathrm{in}$, $\vec{E}_\mathrm{in}$ can be omitted. Using the relationships Eqs. (12), (13) and (16), the formula for reflectance is represented by the following $2\times 2$ matrix equation,

$$(I-r)(I+r)^{-1}=\mu_R^{-1}n=\varepsilon n^{-1}. \tag{22}$$

Using Eq. (22) and the transmittance formula, not shown here but can be derived in a similar way, all matrix components of the optical responses $\varepsilon,\mu$, and $n$ can be obtained in principle. Therefore, the measurement of $r$ only, as in our experiment, is not sufficient to obtain all of them. However, by setting the permeability to $\mu=I$, which reduces to Eq. (4), we can obtain $n$ and $\varepsilon$ from $r$ measured.

### C. Measurement principle of MOKE

Figure 2 shows a schematic optical system for MOKE measurements for the crystal *ca*-surface, where the angles of the polarizers 1 and 2 are set to $\phi_1=90°$ and $\phi_2=-45°$ for measurements, respectively. The electric field of light $\tilde{E}$ (scalar) at angular frequency $\omega$ entering the detector is expressed using the Jones vector/matrix as follows [50–52],

$$\tilde{E} = \begin{pmatrix} \cos\phi_2 \\ \sin\phi_2 \end{pmatrix} \cdot \begin{pmatrix} e^{\frac{i\delta}{2}} & 0 \\ 0 & e^{-\frac{i\delta}{2}} \end{pmatrix} \begin{pmatrix} \tilde{r}_c & -\tilde{\theta} \\ \tilde{\theta} & \tilde{r}_a \end{pmatrix} \begin{pmatrix} \cos\phi_1 \\ \sin\phi_1 \end{pmatrix} \times E_{\mathrm{in}} \exp(-i\omega t) \tag{23}$$

where the retardation $\delta = \delta_0(\omega)\cos(2\pi f_p\, t)$ represents the phase difference between the *x*- and *y*-axes produced by PEM, which oscillates at the frequency $f_p$ of about 50 and 80 kHz, for $SiO_2$ (≳ 0.4 eV) and ZnSe (≲ 0.5 eV) media, respectively. The $\omega$-dependence of $\delta_0$ is calibrated in advance [50]. Note that $\omega$ is for light and $f_p$ is for PEM. The detected signal is proportional to the average intensity (power) of light,

$$I = \frac{|\tilde{E}|^2}{2} \simeq I_{dc} + I_{1f}\cos(1 \times 2\pi f_p t) + I_{2f}\cos(2 \times 2\pi f_p t). \tag{24}$$

The signal with the phase modulation by PEM contains the components of the fundamental frequency $I_{1f}$, the second harmonic $I_{2f}$, and the unmodulated $I_{dc}$, which are extracted by a lock-in amplifier. Using the Bessel functions $J_n \equiv J_n(\delta_0(\omega))$ coming from the expansion coefficients of $\cos\delta$ and $\sin\delta$, and the normalized modulated coefficients $I_1 \equiv I_{1f}/I_{dc}$ and $I_2 \equiv I_{2f}/I_{dc}$, the angles $\theta_{\mathrm{K}|a}$ and $\eta_{\mathrm{K}|a}$ are obtained as,

$$\theta_{\mathrm{K}|a} = \mathrm{Re}\left(\frac{\tilde{\theta}}{\tilde{r}_a}\right) = \frac{r_a\theta + s_a\eta}{R_a} = -\frac{I_2}{4J_2},$$
$$\eta_{\mathrm{K}|a} = \mathrm{Im}\left(\frac{\tilde{\theta}}{\tilde{r}_a}\right) = \frac{r_a\eta - s_a\theta}{R_a} = -\frac{I_1}{4J_1}, \tag{25}$$

where $R_a = r_a^2 + s_a^2$ is the energy reflectivity in the crystal $a$-axis.

### D. Sample

The κ-Cl crystals used were grown from aqueous solution by conventional electrochemical methods. The crystals are platelike and rhombic in shape and approximately 1 mm in diameter and 0.2-0.4 mm thickness. The rhombic surface is the *ca* plane [Fig. 1(b)] as flat as a mirror and can be used as is for optical measurements. We have confirmed that the crystal has good quality based on the optical properties including anisotropy, as well as on the evaluation of resistivity, magnetic susceptibility [Fig. 1(c)], and optical properties of other crystals of the same batch.

### E. Optical measurement

The diagonal reflectivity in the far- and mid-infrared light region and the magneto-optical Kerr effect in the mid-IR light region were performed with the measurement system at BL43IR, SPring-8, Japan [52–55]. Infrared synchrotron radiation enders the measurement system through an FTIR (Bruker, IFS120HR). A Cassegrain mirror with NA = 0.3 is used to focus the light onto a spot diameter of 100-150 mm (near normal incident light with the average angle of incidence ~9°). Therefore, we can use Eq. (4) for analysis. A superconducting magnet equipped can apply a magnetic field of up to ±14 T perpendicular to the sample plane (polar Kerr configuration).

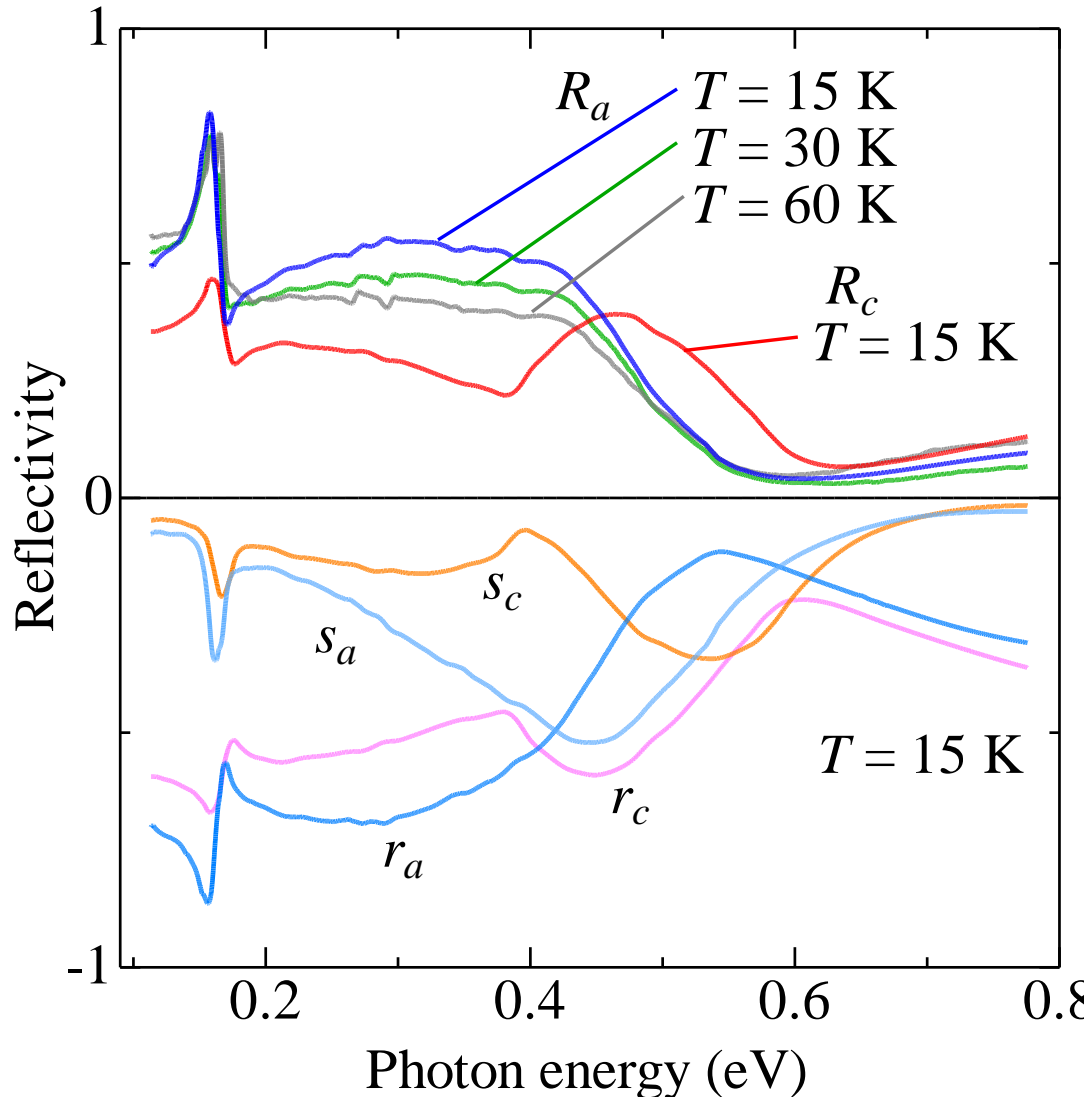

FIG 3. Spectra of energy reflectivity $R_a$ at 60, 30, and 15 K in $E||a$ and $R_c$ at 15 K in $E||c$. The real $r$ and imaginary $s$ part of the complex amplitude reflectance (Fresnel coefficient) are shown in $E||a$ and $E||c$ at 15 K. Note that both $r$ and $s$ are negative and the relation of $R_\alpha = r_\alpha^2 + s_\alpha^2$ $(\alpha = c, a)$.

As schematically illustrated in Fig. 2, the incident light was fixed vertically (∥ $a$) for the MOKE measurements since the synchrotron radiation light source is elliptically polarized with the vertical direction stronger, and the incident angle to the sample must be adjusted within about 0.2°. The crystal axes $(\alpha, \beta, \gamma)$ in subsection II.B correspond to the actual κ-Cl crystal axes $(c, a, b)$, respectively. The κ-Cl crystals were held in place by gold wires to make thermal contact and to avoid the pressurization effect of thermal contraction of the metal stage in a cryostat. The magnetic field is applied along the *b*-axis. The Faraday effect (nearly constant a few mrad/T for rotation and almost zero for ellipticity) due to the KBr window of the cryostat were subtracted. For the $BaF_2$ window the Faraday effects were negligible. We used two types of PEM (Tokyo Instruments): ZnSe for ≲ 0.5 eV and $SiO_2$ for ≳ 0.4 eV.

For the near-infrared, visible, and ultraviolet regions, measurements were performed in the laboratory. After obtaining a set of spectra, the complex diagonal reflectance was calculated by Kramers-Kronig transformation (KKT).

## III. RESULTS

### A. Diagonal reflectivity

Figure 3 shows the diagonal reflectivity spectra in $E \parallel a$ at $T = 60, 30$, and 15 K and $E \parallel c$ at 15 K. The energy reflectivity $R_a$ at around 0.15-0.40 eV monotonically increases with decreasing temperature. Both $R_a$ and $R_c$ show a sharp resonance at 0.16-0.18 eV due to the molecular vibration. A large dip structure in $R_c$ at 0.4 eV corresponds to the change in the π-electron transition from intra- to inter-dimer transition with increasing photon energy [45,46]. The complex reflectivity coefficients $r_\alpha + is_\alpha$ at 15 K were obtained by the KKT with the spectra from about 0.03 to 3 eV, where the spectra outside of the energy region shown in Fig. 3 do not show remarkable structures. The real and imaginary parts of the reflection coefficients are negative since the time dependence of the electric field is defined as $\exp(-i\omega t)$ . These results, including the complex coefficients, are consistent well with the precedent report [45].

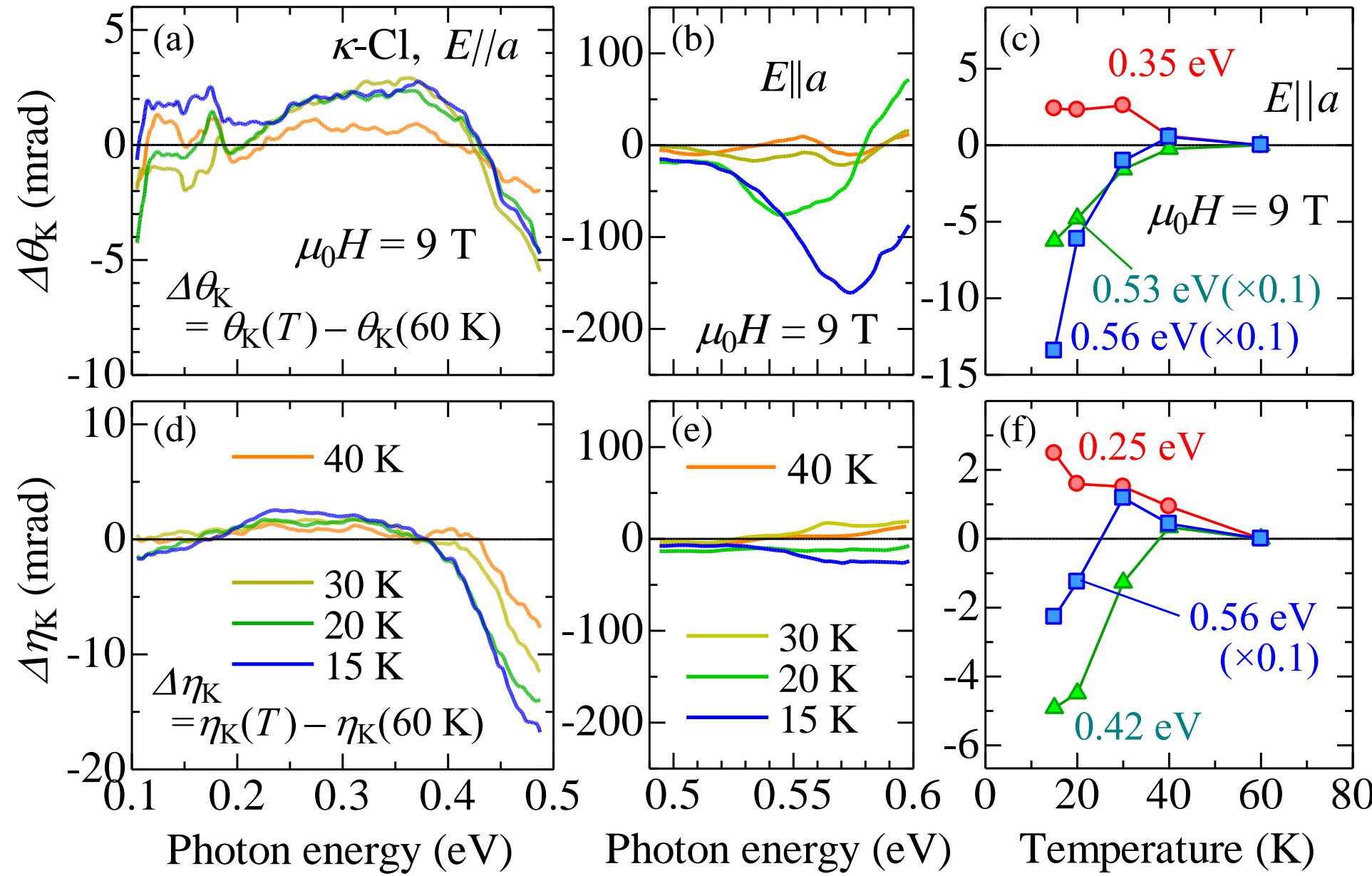

FIG 4. Temperature dependence of MOKE spectra in E||a at 9 T. To remove the unwanted effects of windows and mirrors, the spectrum at temperature below 60K is subtracted by the spectra at 9 T and 60 K. The upper panels show the Kerr rotation spectra in the (a) low and (b) high energy region, and (c) the temperature dependence at 0.35, 0.53, and 0.56 eV. The lower panels show the Kerr ellipticity spectra in the (d) low and (e) high energy region, and the temperature dependence of at 0.25, 0.42, and 0.56 eV. In (c) and (f), 0.1 is multiplied due to large angles in the high energy region.

### B. Temperature dependence

Figure 4 shows the MOKE spectra of angles $\theta_\mathrm{K}$ and $\eta_\mathrm{K}$ at 9 T in $E \parallel a$. The spectra in Figs. 4(a) and 4(b) are the difference from $\theta_\mathrm{K}$ at 60 K, similarly Figs. 4(d) and 4(e) are the difference from $\eta_\mathrm{K}$ at 60 K in the low and high energy regions, respectively. In the low energy region, $\theta_\mathrm{K}$ and $\eta_\mathrm{K}$ are small but all the spectra enhance below about $T_\mathrm{N}$. In the high energy region [Fig. 4(b)], $\Delta\theta_\mathrm{K}$ increases significantly in the negative direction down to 15 K, reaching about -150 mrad (-8.6°). This is due to the very low reflectivity ($<$ 10 %) as shown in Fig. 3. Figures 4(c) and 4(f) show the temperature dependence of the angles at several energies, all of which become apparent below around $T_\mathrm{N}$.

At magnetic fields below 5-6 T, the MOKE signals were unclear, even below $T_\mathrm{N}$, whereas the MOKE signal at 9 T seems to occur above $T_\mathrm{N}$ as shown in Figs. 4. We consider that this is a kind of critical behavior under a strong magnetic field, at temperatures where the spin-spin interaction is significant due to the weak anisotropy in the spin direction, as follows.

In Fig. 1(c), magnetic susceptibility decreases gradually below 40-50 K, indicating that the magnetic interaction begins to be significant with respect to temperature. Reference [56] also clarified by NMR measurements at 7.4 T, that antiferromagnetic spin correlations begin to emerge in this temperature range. At 7 T, a spin tilt of 10% is observed in the *M-H* curve in the AF phase [26]. The anisotropy of single dimer 1/2-spin is weak, and 9 T should be sufficient to eliminate the cancellation effect between the two layers in κ-Cl. Therefore, the appearance of MOKE above $T_\mathrm{N}$ is attributed to the spins tilting along the 9 T magnetic field and beginning to correlate antiferromagnetically in this temperature regime.

### C. Field dependence

To clarify whether the MOKE signal is proportional to the net magnetic moment or not, we show the field dependence of the spectra of $\theta_\mathrm{K}$ and $\eta_\mathrm{K}$ at 15 K in Fig. 5, where the angles are calculated as $\theta_\mathrm{K}(H) = [\theta_\mathrm{K}(+H) - \theta_\mathrm{K}(-H)]/2$, as well as for $\eta_\mathrm{K}$. At lower magnetic field below 6-7 T, all the spectra in Fig. 5 are small, while with increasing field the spectra non-monotonically enhance with some structures. In Fig. 5(a), at 0.12-0.15 eV a sharp peak grows with magnetic field and spectra go negatively toward higher energy. In the higher energy region shown in Fig. 5(b), $\theta_\mathrm{K}$ becomes the lowest at

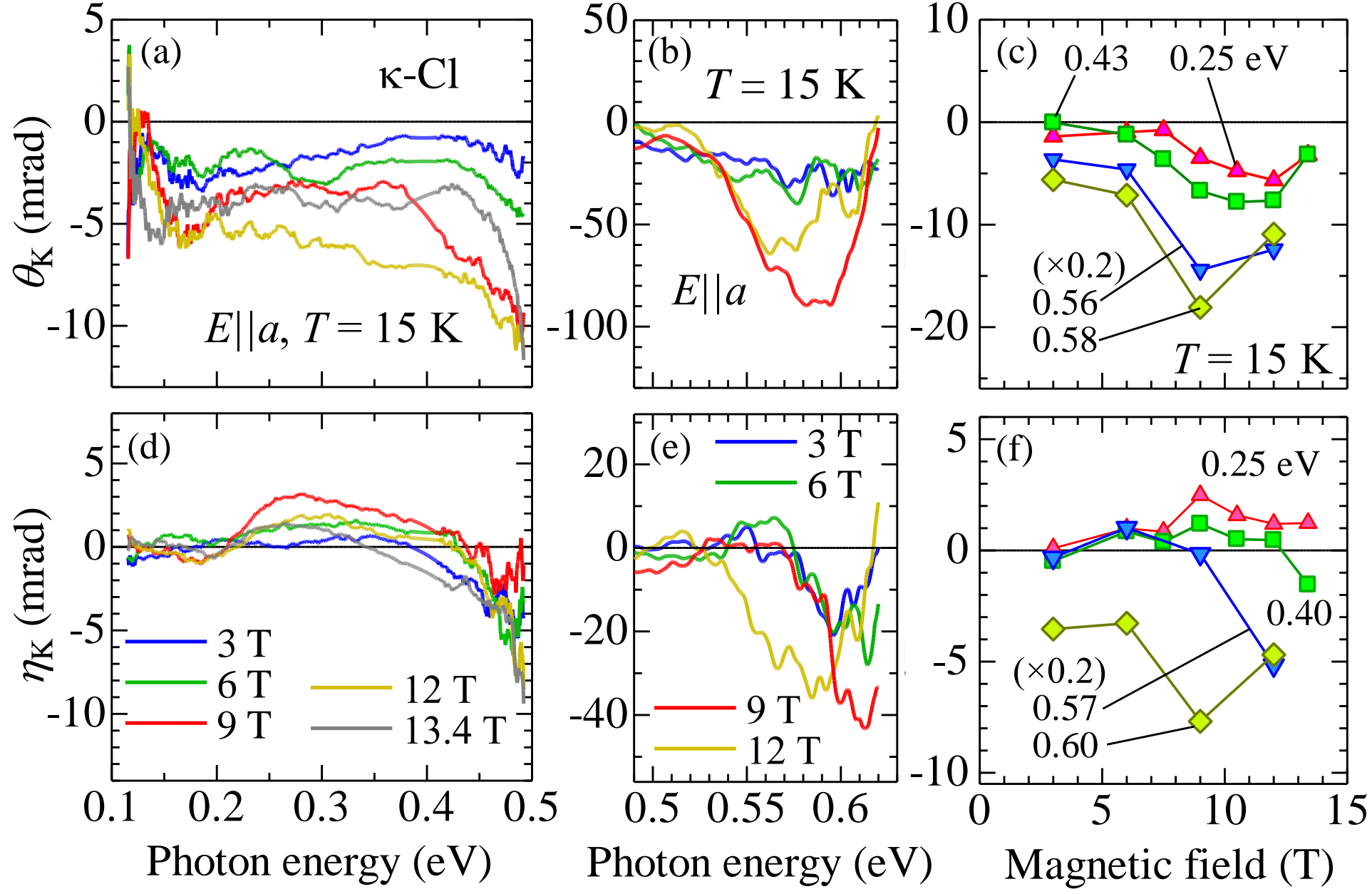

FIG 5. Magnetic field dependence of MOKE spectra in E||a at 15 K, where the spectra are the odd component of the magnetic field, and the corresponding spectrum at 60 K is subtracted. The upper panels show the Kerr rotation spectra in the (a) low and (b) high energy region, and (c) the field dependence at 0.25, 0.43, 0.56, and 0.58 eV. The lower panels show the Kerr ellipticity spectra in the (d) low and (e) high energy region, and the field dependence at 0.25, 0.35, 0.40, 0.57, and 0.60 eV. In (c) and (f), 0.2 is multiplied due to large angles in the high energy region.

9 T, then increases at 12 T. The peak structure, as well as the peak in Fig. 5(a), seems to shift to lower energy with increasing magnetic field. In Fig. 5(d), non-monotonic increase and decrease in $\eta_\mathrm{K}$ is observed, where $\eta_\mathrm{K}$ becomes the largest at 9 T in the energy region of 0.2-0.4 eV. As well as $\theta_\mathrm{K}$, in Fig. 5(e) the peak of $\eta_\mathrm{K}$ shifts toward the low energy. These nonlinear field dependencies are clearly seen also in Figs. 5(c) and 5(f). In Fig. 5(c) $\theta_\mathrm{K}$ is almost zero in magnetic fields below 5-6 T, begins to increase sharply in the negative direction at 6-7 T, saturates around 9 T, and moves toward zero. The same behavior can be seen in $\eta_\mathrm{K}$. Since the net magnetic moment of the canted-AF state is proportional to the magnetic field after the spin flop [26], the observed MOKE signals have no direct relationship with the canted magnetization ($\Delta M_b$) or magnetic field, which is qualitatively consistent with the theoretical results for MOKE/AHE due to altermagnetism. In addition, the peak energy shifts without phase transitions under a magnetic field of ~10 T is unusual.

### D. Diagonal and Off-diagonal elements

For a more detailed discussion, $\tilde{\theta}_\mathrm{K}$, $\tilde{\theta}$, diagonal, and off-diagonal conductivity $\tilde{\sigma}_{ca}$ spectra at 15 K, 9 T, $E \parallel a$ are shown in Fig. 6, two of which are scaled for clarity. The diagonal conductivities $\sigma_a$ in $E \parallel a$ and $\sigma_c$ in $E \parallel c$ were calculated using the spectra shown in Fig. 3. In particular, $\theta_\mathrm{K}$ seems to be antisymmetric as the center around 0.4 eV. This trend is the same for $\tilde{\theta}$ and $\tilde{\sigma}_{ca}$. The significant difference of the peak values at both ends in $\theta_{\mathrm{K}|a}$ become small in $\tilde{\theta}$, and become comparable in $\tilde{\sigma}_{ca}$.

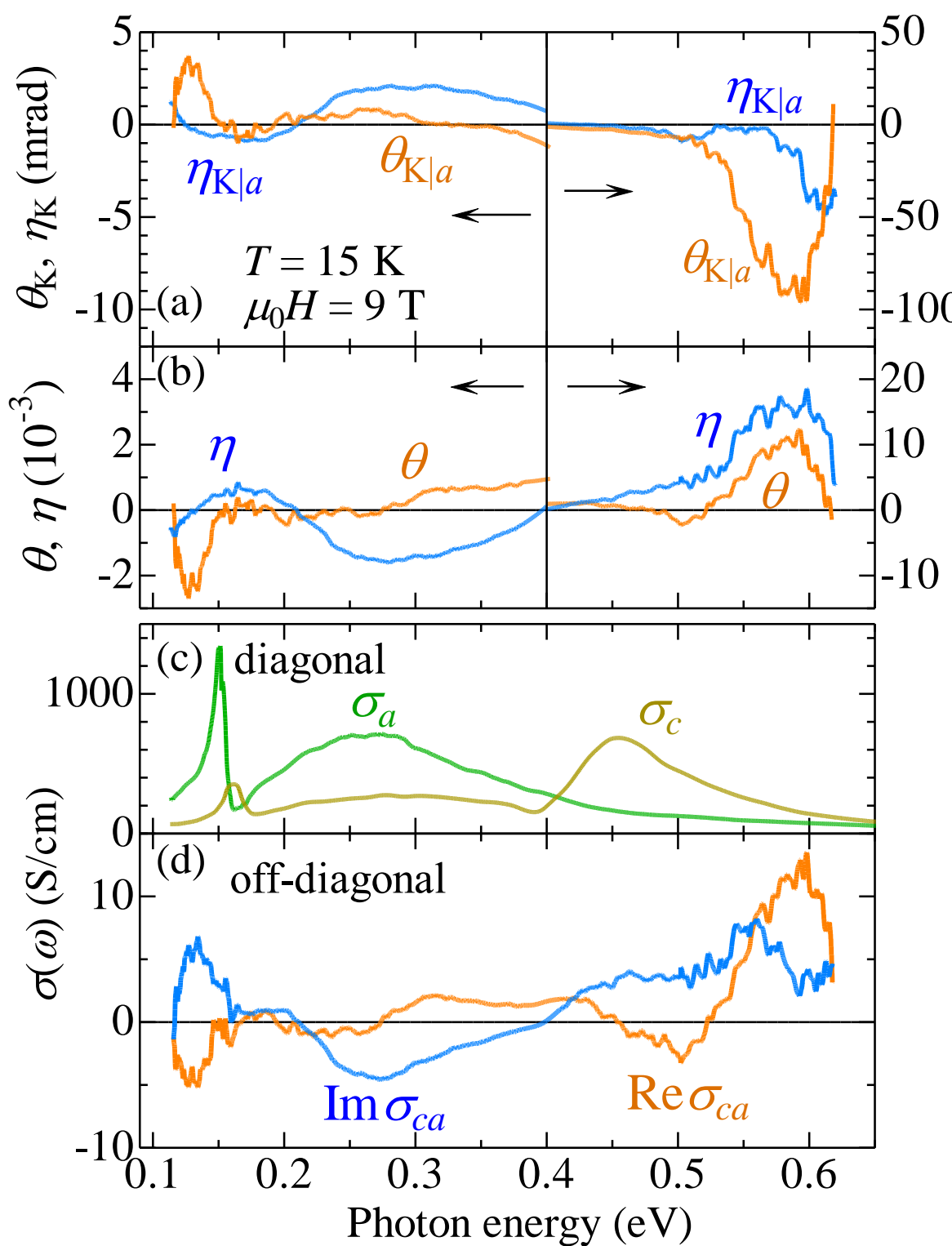

FIG 6. Spectra of (a) Kerr rotation $\theta_{\mathrm{K}|a}$ and ellipticity $\eta_{\mathrm{K}|a}$ in $E||a$, (b) the real $\theta$ and imaginary $\eta$ part of off-diagonal amplitude reflectivity, (c) diagonal and (d) off-diagonal optical conductivity at 9 T and 15 K.

$\tilde{\sigma}_{ca}$ is calculated using diagonal elements according to Eq. 4, which includes the sum of complicated diagonal terms. However, the resulting spectra can be easily interpreted due to the simple band structure in organic systems.

First, comparing the off-diagonal $\tilde{\sigma}_{ca}$ with the diagonal $\sigma_a$ and $\sigma_c$ in the central spectra (~0.2-0.5 eV), $\sigma_a$ and $\sigma_c$ roughly corresponds to the imaginary and real part of $\tilde{\sigma}_{ca}$, respectively, i.e. $\tilde{\sigma}_{ca} \propto -\sigma_c - i\sigma_a$. This correspondence is not trivial, although Eq. 4 includes the diagonal conductivities. The spectral features of $\sigma_a$ and $\sigma_c$ are explained by two transitions of the $\pi$-electron within a dimer and between adjacent dimers along the $c$-direction, respectively [9,45,57]. As is discussed in the next section, these correspondences can be understood by the theoretical result that the $\pm k_x k_y$-type band structure of the altermagnet derives from normal electron hoppings rather than SOC.

On the other hand, second, the spectra around the ends (0.10-0.15 and 0.5-0.6 eV) cannot be attributed to the diagonal components. Such band-edge peaks are typically seen in ferromagnets attributed to the up- and down-spin band difference [44]. Therefore, the end-peaks clearly correspond with the AF spin ordering. Note that the sharp spectra seen around 0.15 eV in diagonal conductivities are due to molecular vibrations whose effect on $\tilde{\sigma}_{ca}$ is small because of almost zero $\tilde{\theta}$. The width of the end-peaks (~0.05 and 0.1 eV at low and high energy, respectively) corresponds to the split width of the up- and down-spin bands. This splitting energy cannot be explained only by the small spin-orbit coupling (≲1 meV) in the organic molecular based κ-Cl [9,58,59]. Both of these features amplify the $\tilde{\sigma}_{ca}$ of κ-Cl to the extent that it is comparable to that of ferrites [60].

## IV. DISCUSSION

We discuss the attributions of the observed $\tilde{\sigma}_{ca}$ in relation to the altermagnetism and others. The symmetry to produce the $s_c k_c k_a$-type spin-band splitting in κ-Cl is intrinsically the same as that to produce the piezomagnetic effect [29,61,62] and symmetric spin current [13], as described in Introduction. These effects can occur at once. In the piezomagnetic material, the crystal distortion produces the magnetization. This effect inversely means that the crystal distortion occurs antisymmetrically by the reversal of applied magnetic field or antiferromagnetic spins as shown in Fig. 7. Therefore, even if a field-odd component $\sigma_{xy}(-H) = -\sigma_{xy}(H)$ is measured, we cannot distinguish whether the signal stems from the antisymmetric AHE/MOKE or symmetric inverse-piezomagnetic distortion. The MOKE can detect the signals from several related origins [42], including the field-odd and symmetric $\sigma_{xy}(H) = -\sigma_{xy}(-H) = \sigma_{yx}(H)$ component [38,43]. However, we can distinguish them in $\tilde{\sigma}_{ca}(\omega)$ by considering the diagonal-term dependencies, as follows.

As shown in Fig. 7, rotating the crystal by a small angle $\Delta\varphi$ adds $(\tilde{\sigma}_c - \tilde{\sigma}_a)\Delta\varphi$ to the off-diagonal conductivities. Similarly, a lattice distortion, e.g., from a rectangle to a parallelogram, in general, appends a linear combination of the diagonal elements to $\tilde{\sigma}_{ca}$ as $\tilde{\sigma}_c\Delta\varphi_1 + \tilde{\sigma}_a\Delta\varphi_2$, where the coefficients ($\Delta\varphi_1$ and $\Delta\varphi_2$) must be real numbers because the distortion is static and the angle is real. Since $\sigma_{ca}$ in Fig. 6(d) is field-odd component and the diagonal conductivities are field-even, the angle $\Delta\varphi$ should be field-odd, $\Delta\varphi(-H) = -\Delta\varphi(H)$. Therefore, the inverse-piezomagnetic distortion mechanism is a possible explanation for the mid-region spectra of $\mathrm{Re}\{\tilde{\sigma}_{ca}\} \sim -\sigma_c\Delta\varphi$ (~0.4-0.55 eV), where $\Delta\varphi$ is estimated to be about 0.01 rad (0.6°). The off-diagonal term due to the static distortion should be a symmetric tensor.

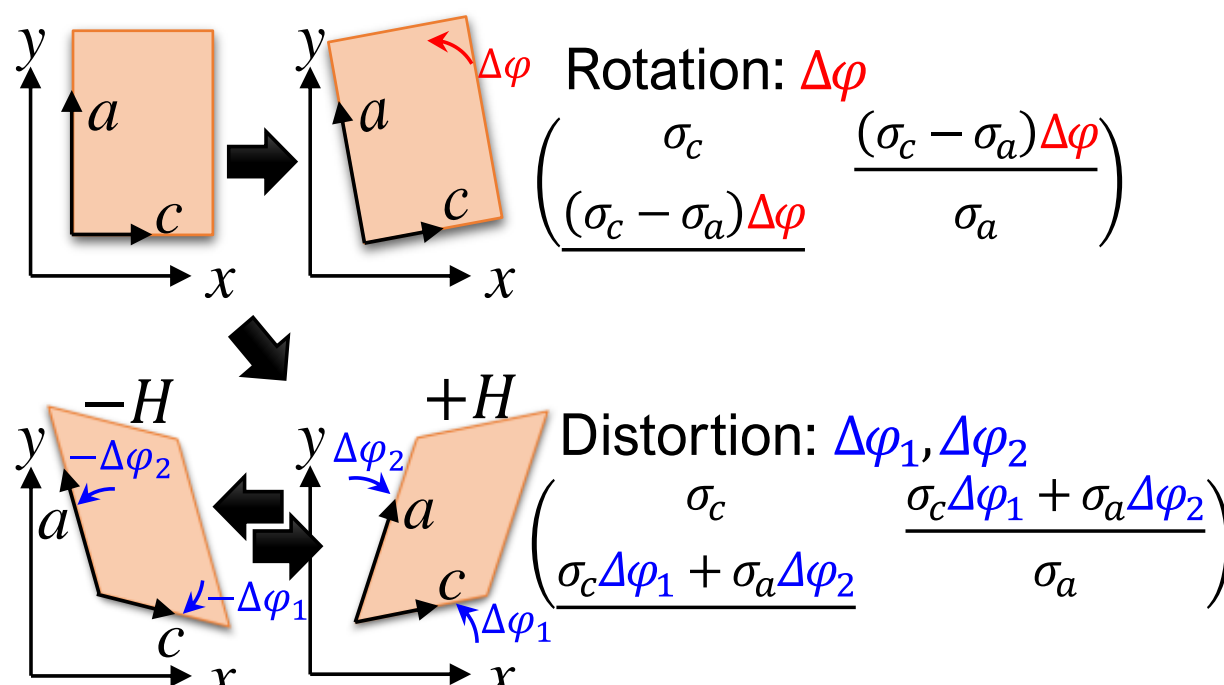

FIG 7. Effects of crystal rotation and distortions on the off-diagonal matrix elements. The terms underlined in the matrices are added by the effect.

On the other hand, the imaginary part of the mid-region spectra $\mathrm{Im}\{\tilde{\sigma}_{ca}\} \propto -\sigma_a$ (~0.2-0.4 eV) is qualitatively different from the real part because $j_c$ is proportional to $j_a$, i.e., $j_c = \tilde{\sigma}_{ca}E_a \propto -i\sigma_a E_a = -ij_a$ for applied $E_a$ or $j_a$. Considering the spin current generation, the importance is that spin and flow can be reversed simultaneously as $s_c j_c = (-s_c)(-j_c)$. However, in the imaginary part, the sign of $j_c$ is fixed by $j_a(\propto ij_a)$, so the opposite spin current of $(-s_c)(-j_c)$ cannot be generated. In other words, the imaginary part in this region implies current rotation without spin currents and a standard antisymmetric field-odd tensor component. This contribution to $\tilde{\sigma}_{ca}$ may correspond to the MOKE/AHE in the strong dimerization limit in Ref. [9], where the MOKE/AHE remains nonzero, although the $sk_x k_y$-type spin band splitting has disappeared.

In addition to the end peaks already mentioned, we found that the observed $\tilde{\sigma}_{ca}$ spectrum has three features. These are: the end peaks due to spin band splitting, which is independent of the diagonal conductivities; the real part in the middle region, which is expected to be associated with the inverse piezomagnetic effect; and the imaginary part, which produces the standard antisymmetric rotating current without spin current. We emphasize that the appropriate measurement and the correct calculation of $\tilde{\sigma}_{ca}$ are quite important.

Next, we discuss the tensor symmetry of off-diagonal elements, i.e., $\sigma_{ca}$ and $\sigma_{ac}$. Note that in our MOKE measurement method for anisotropic magnets, different from the methods using circularly polarized lights, $\tilde{\sigma}_{ca}$ can be obtained independently from the other off-diagonal element $\tilde{\sigma}_{ac}$. All the *xy*-elements of the matrices are proportional to each other $r_{xy} \propto n_{xy} \propto \varepsilon_{xy} \propto \sigma_{xy}$ but have mathematically no relation to the *yx*-elements.

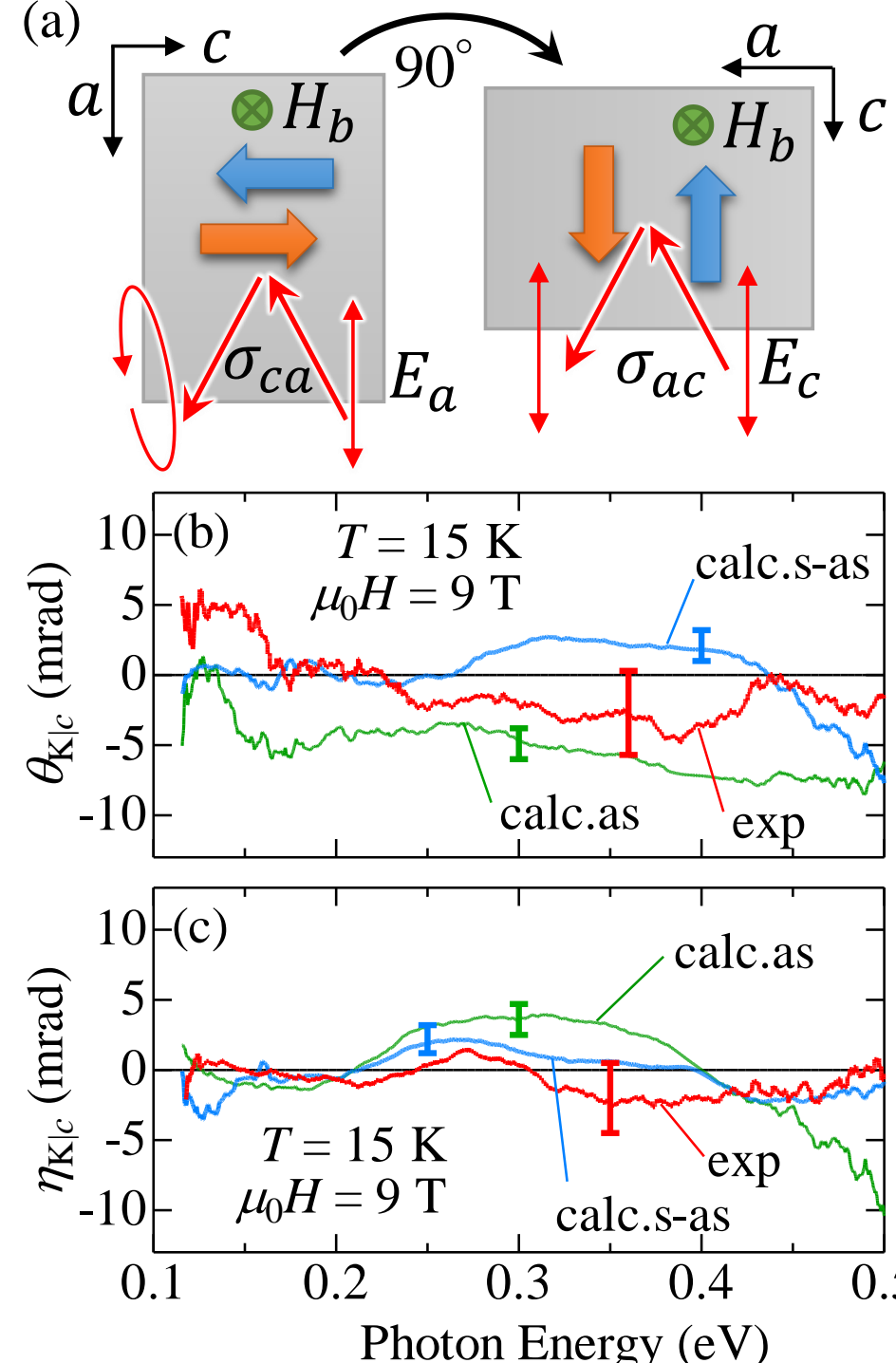

FIG 8. (a) Schematics of MOKE measurements in $E \parallel a$ for $\sigma_{ca}$ and $E \parallel c$ for $\sigma_{ac}$. Not the incident polarization direction but the sample orientation was changed because of the efficient use of elliptically polarized synchrotron radiation. Experimental (exp.) and calculated (calc.s-as and calc.as) spectra of the Kerr rotation (b) and ellipticity (c) in $E \parallel c$. Error bars are also shown. The calculations were performed using the results of $E \parallel a$, with the symmetric real part and antisymmetric imaginary part $\tilde{\sigma}_{ac} = \mathrm{Re}\{\tilde{\sigma}_{ca}\} - i\,\mathrm{Im}\{\tilde{\sigma}_{ca}\}$ for calc.s-as, and $\tilde{\sigma}_{ac} = -\tilde{\sigma}_{ca}$ for calc.as.

We tried to observe the off-diagonal amplitude reflectivity $\tilde{r}_{ac} \propto \tilde{\sigma}_{ac}$ in $E \parallel c$ and $H \parallel b$ several times in the same optical settings by rotating crystal as illustrated in Fig. 8(a). However, as shown in Figs. 8(b) and 8(c), the observed spectra are close to zero within the error. The errors of the spectra in $E||c$ are about twice compared with that of $E||a$, mainly because $R_c$ is about a half of $R_a$ (see Fig. 3). In Figs. 8(b) and 8(c), we also show the expected spectra with the symmetric real and antisymmetric imaginary parts of $\tilde{\sigma}_{ac} = \mathrm{Re}\{\tilde{\sigma}_{ca}\} - i\mathrm{Im}\{\tilde{\sigma}_{ca}\}$ (calc.s-as) as discussed above and the antisymmetric $\tilde{\sigma}_{ac} = -\tilde{\sigma}_{ca}$ (calc.as) for comparison. We cannot consider $\theta_{K|c}$ as a significant signal due to the large error and noise, partly due to the Faraday rotation of the cryostat window (Sect. II.E). On the other hand, $\eta_{K|c}$ is closer to the calc.s-as spectrum than to the calc.as one at about 0.15-0.5 eV, consistent with the discussion above. However, the experimental $\eta_{K|c}$ does not show the end-peak structure at 0.12-0.15 eV which is seen in the calc.s-as spectrum but not in the calc.as. Based on the results in $E \parallel a$ and $E \parallel c$, the off-diagonal elements can be composed of the symmetric and antisymmetric terms in the mid-spectral region, but not at the spectral end. The mid-region spectra of $\tilde{\sigma}_{ac}$ and $\tilde{\sigma}_{ca}$ are proportional to the diagonal conductivities, while the end peaks are independent of the diagonal conductivities. This qualitative difference indicates that the MOKE in the two regions have different origins or symmetries.

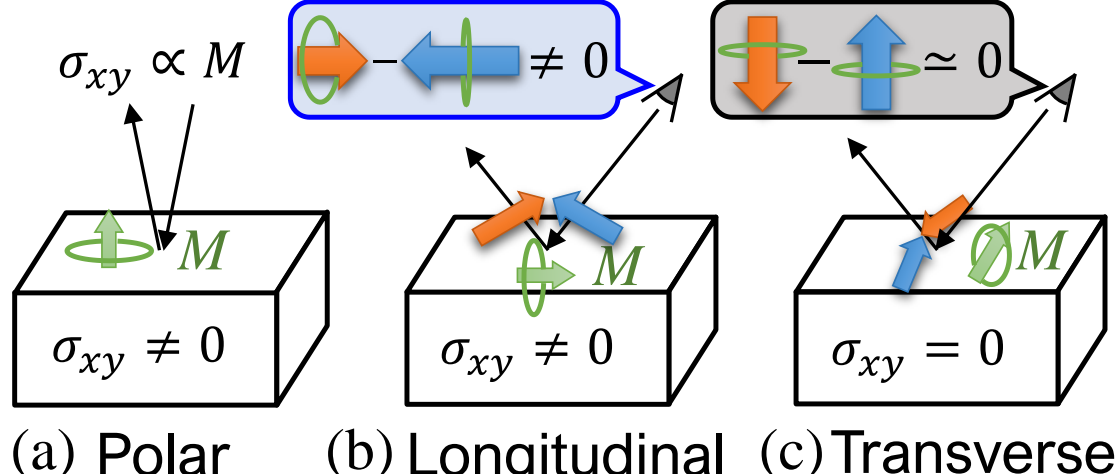

FIG 9. Schematics of the configurations of (a) polar, (b) longitudinal, and (c) transverse MOKE corresponding to the spectra in the middle region of $\tilde{\sigma}_{ca}$ and $\tilde{\sigma}_{ac}$, the edges of $\tilde{\sigma}_{ca}$, and $\tilde{\sigma}_{ac}$, respectively. The appearance of $\sigma_{xy}$ corresponds to the appearance of the disk perpendicular to the magnetization vector $M$ in *ferromagnet*.

As noted by McClarty and Rau [20], AHE and MOKE have a symmetry that transforms like the magnetization. Typically, AHE and MOKE are proportional to the magnetization. In addition, the Néel order parameter appears in altermagnets, and the spin-orbit interaction connects the Néel order parameter to the net magnetization. Consequently, the AHE and MOKE with the same symmetry as the net magnetization become proportional to the magnitude of the Néel order parameter. However, MOKE is an optical phenomenon which can also exhibit other responses. These responses include excitations where the spins transit between the sublattice bands ($\Delta s_c = \pm 1$) via SOC as the absorption difference of the circularly polarized lights, MCD, called the paramagnetic term [44,60]. The end peaks correspond to this excitation and the width of the peak, the difference in the spin bands, is large in agreement with the theoretical calculation.

This optical transition, which appears as the end peak, is directly related to the Néel vector or AF spin direction along the *c*-axis, independent of the spin-canting magnetization or charge/spin hopping resulting in the dc AHE. In other words, there is a directional difference between the two responses; one is due to the electron hopping in the altermagnetic $xy(ca)$ plane, and the other is due to the change in the $\pm c$-spin direction by circularly polarized light. The latter origin should be unique to the ac MOKE and not to the dc AHE. The circularly polarized light should have a component along the *c*-axis. Therefore, the effect of oblique incidence must be considered.

MOKE is roughly proportional to the inner product between the ray vector (with oblique incidence angle $\phi$) and the magnetization $M$ in *ferromagnet*. In κ-Cl, both of the net magnetization $\Delta M_b$ and the Néel vector $N_c$ (or spin) can correspond to $M$. Figure 9 shows the (a) polar, (b) longitudinal, and (c) transverse Kerr configurations, corresponding to the spectra in the middle region, the edge peaks of $\sigma_{ca}$, and $\sigma_{ac}$, respectively. Considering their inner product, the MOKE signal is proportional to $\cos\phi$ in polar, $\sin\phi$ in longitudinal, and zero in transverse Kerr configurations. The crystal rotation in Fig. 8(a) has little effect on the spectra in the middle region, but a large effect on the end peaks ($\propto \sin\phi$ or zero). Note, however, that all the responses follow $N_c$ and that significant spin canting is required for the large difference between the end peaks in Fig. 9(b) and 9(c).

In connection with the oblique angle ~9° (see Fig. 2 and Sect. II. E), we discuss quantitative angles and the nonlinear field dependence of the MOKE. We can compare experimental and theoretical results by taking the ratio of off-diagonal to diagonal conductivities, $\sigma_{ca}/\sigma_a$, in the middle region of $\tilde{\sigma}_{ca}$. The ratio in the theoretical calculation [9] is about $10^{-3}$, but in our experiments both the real and imaginary parts are 10 times larger ($10^{-2}$ ~ 0.6°). These results and the nonlinearity in Fig. 5 suggest that the experimental environment is in the nonlinear regime [20] with respect to the spin canting and over the range of spin structures where theoretical calculations [9] have been performed. In fact, the *M-H* curve [26] shows a spin tilt of about 10% (about 6°) at 7 T, which can be estimated to be more than 10° at 12-13 T. Since $\sigma_{xy}$ of the altermagnet follows the magnitude of the Néel vector, the decrease in MOKE at high fields can be attributed to the reduction in $|s_c|$ due to the large spin tilt. At about 6 T, the Néel order parameter entered the nonlinear region, bringing the MOKE signal into our observable range. The incident angle of 9°, as shown in Figs. 2 and 9, is much larger than the MOKE in the middle region ($< 10$ mrad $= 0.6°$) and the inverse piezomagnetic distortion ($\sim 0.6°$). However, it is comparable to the end peak of ~100 mrad = 6° on the high energy side and the spin tilt angle (~6-10° at 7-12 T).

## V. CONCLUSIONS

In summary, we have revealed the off-diagonal optical response in the organic Mott insulator κ-Cl by the infrared MOKE spectra below 60 K. The *ca* plane MOKE spectra appear near the antiferromagnetic transition temperature and show the nonlinear magnetic field dependence different from that of the weak-ferromagnetic magnetization, which is qualitatively consistent with the theoretically suggested properties of the altermagnetism. Using the general formulae for reflection and refractive index, the off-diagonal optical conductivity of the entire π-electron transition band was determined. We found that the off-diagonal conductivity spectra $\tilde{\sigma}_{ca}$ have three different characters: the real, imaginary parts in the mid-spectral region, and the peaks at the spectral ends. The spectra in the middle region are nearly proportional to the anisotropic diagonal conductivities as $\tilde{\sigma}_{ca} \propto -\tilde{\sigma}_c - i\tilde{\sigma}_a$, where the real and imaginary parts can be interpreted as the symmetric inverse piezomagnetic effect and antisymmetric rotational current, respectively. The end peaks show the large energy width and intensity, which suggests the spin-band separation of as much as 0.05-0.1 eV for the organic crystals with small spin-orbit interactions. The asymmetry of $\tilde{\sigma}_{ac}$ can be explained by the symmetric real and antisymmetric imaginary parts of $\tilde{\sigma}_{ca}$ in the middle region. The asymmetric end peaks can be interpreted as the spin-reversal optical transition directly related to the Néel vector in the *c*-direction. The relationships between the three MOKE configurations and the symmetry of the off-diagonal conductivities are discussed. These results suggest the altermagnetic MOKE response of κ-Cl.

## ACKNOWLEDGMENTS

We would like to thank M. Naka and H. Seo for their fruitful discussion. This research was supported by JSPS KAKENHI Grants Numbers 23H04015, 23K25811, 23K22420, 23K17659, 23K03271, 23H01114, and 22H01149. The synchrotron radiation experiments were performed at the BL43IR of SPring-8 with the approval of the Japan Synchrotron Radiation Research Institute (JASRI) (Proposal Numbers 2024B1236, 2024A1193, 2023B1489, 2023B1397, 2023A1462, 2023A1229, 2022B1514, and 2016A0073).

---

[1] N. Nagaosa, J. Sinova, S. Onoda, A. H. MacDonald, and N. P. Ong, Anomalous Hall effect, Rev. Mod. Phys. **82**, 1539 (2010).

[2] D. Xiao, M.-C. Chang, and Q. Niu, Berry phase effects on electronic properties, Rev. Mod. Phys. **82**, 1959 (2010).

[3] N. Nagaosa and Y. Tokura, Emergent electromagnetism in solids, Phys. Scr. **T146**, 014020 (2012).

[4] Y. Machida, S. Nakatsuji, S. Onoda, T. Tayama, and T. Sakakibara, Time-reversal symmetry breaking and spontaneous Hall effect without magnetic dipole order, Nature **463**, 210 (2010).

[5] H. Chen, Q. Niu, and A. H. MacDonald, Anomalous Hall Effect Arising from Noncollinear Antiferromagnetism, Phys. Rev. Lett. **112**, 017205 (2014).

[6] S. Nakatsuji, N. Kiyohara, and T. Higo, Large anomalous Hall effect in a non-collinear antiferromagnet at room temperature, Nature **527**, 212 (2015).

[7] L. Šmejkal, A. H. MacDonald, J. Sinova, S. Nakatsuji, and T. Jungwirth, Anomalous Hall antiferromagnets, Nat Rev Mater **7**, 482 (2022).

[8] H. Takagi et al., Spontaneous topological Hall effect induced by non-coplanar antiferromagnetic order in intercalated van der Waals materials, Nat. Phys. **19**, 961 (2023).

[9] M. Naka, S. Hayami, H. Kusunose, Y. Yanagi, Y. Motome, and H. Seo, Anomalous Hall effect in κ-type organic antiferromagnets, Phys. Rev. B **102**, 075112 (2020).

[10] L. Šmejkal, R. González-Hernández, T. Jungwirth, and J. Sinova, Crystal time-reversal symmetry breaking and spontaneous Hall effect in collinear antiferromagnets, Sci. Adv. **6**, eaaz8809 (2020).

[11] L. Šmejkal, J. Sinova, and T. Jungwirth, Emerging Research Landscape of Altermagnetism, Phys. Rev. X **12**, 040501 (2022).

[12] I. Mazin and The PRX Editors, Editorial: Altermagnetism—A New Punch Line of Fundamental Magnetism, Phys. Rev. X **12**, 040002 (2022).

[13] M. Naka, S. Hayami, H. Kusunose, Y. Yanagi, Y. Motome, and H. Seo, Spin current generation in organic antiferromagnets, Nat Commun **10**, 4305 (2019).

[14] R. González-Hernández, L. Šmejkal, K. Výborný, Y. Yahagi, J. Sinova, T. Jungwirth, and J. Železný, Efficient Electrical Spin Splitter Based on Nonrelativistic Collinear Antiferromagnetism, Phys. Rev. Lett. **126**, 127701 (2021).

[15] T. Aoyama and K. Ohgushi, Piezomagnetic properties in altermagnetic MnTe, Phys. Rev. Materials **8**, L041402 (2024).

[16] H.-Y. Ma, M. Hu, N. Li, J. Liu, W. Yao, J.-F. Jia, and J. Liu, Multifunctional antiferromagnetic materials with giant piezomagnetism and noncollinear spin current, Nat Commun **12**, 2846 (2021).

[17] L. Šmejkal, J. Sinova, and T. Jungwirth, Beyond Conventional Ferromagnetism and Antiferromagnetism: A Phase with Nonrelativistic Spin and Crystal Rotation Symmetry, Phys. Rev. X **12**, 031042 (2022).

[18] Y. Taguchi, Y. Oohara, H. Yoshizawa, N. Nagaosa, and Y. Tokura, Spin Chirality, Berry Phase, and Anomalous Hall Effect in a Frustrated Ferromagnet, Science **291**, 2573 (2001).

[19] K. Ohgushi, S. Murakami, and N. Nagaosa, Spin anisotropy and quantum Hall effect in the *kagomé* lattice: Chiral spin state based on a ferromagnet, Phys. Rev. B **62**, R6065 (2000).

[20] P. A. McClarty and J. G. Rau, Landau Theory of Altermagnetism, Phys. Rev. Lett. **132**, 176702 (2024).

[21] A. Hariki et al., X-Ray Magnetic Circular Dichroism in Altermagnetic α-MnTe, Phys. Rev. Lett. **132**, 176701 (2024).

[22] A. Hariki, Y. Takahashi, and J. Kuneš, X-ray magnetic circular dichroism in $RuO_2$, Phys. Rev. B **109**, 094413 (2024).

[23] U. Geiser et al., Strain index, lattice softness and superconductivity of organic donor-molecule salts, Physica C: Superconductivity **174**, 475 (1991).

[24] U. Welp, S. Fleshler, W. K. Kwok, G. W. Crabtree, K. D. Carlson, H. H. Wang, U. Geiser, J. M. Williams, and V. M. Hitsman, Weak ferromagnetism in κ-$(ET)_2Cu[N(CN)_2]Cl$, where (ET) is bis(ethylenedithio)tetrathiafulvalene, Phys. Rev. Lett. **69**, 840 (1992).

[25] K. Miyagawa, A. Kawamoto, Y. Nakazawa, and K. Kanoda, Antiferromagnetic Ordering and Spin Structure in the Organic Conductor, κ-$(BEDT\text{-}TTF)_2Cu[N(CN)_2]Cl$, Phys. Rev. Lett. **75**, 1174 (1995).

[26] R. Ishikawa, H. Tsunakawa, K. Oinuma, S. Michimura, H. Taniguchi, K. Satoh, Y. Ishii, and H. Okamoto, Zero-Field Spin Structure and Spin Reorientations in Layered Organic Antiferromagnet, κ-$(BEDT\text{-}TTF)_2Cu[N(CN)_2]Cl$, with Dzyaloshinskii–Moriya Interaction, J. Phys. Soc. Jpn. **87**, 064701 (2018).

[27] S. Hayami, Y. Yanagi, M. Naka, H. Seo, Y. Motome, and H. Kusunose, *Multipole Description of Emergent Spin–Orbit Interaction in Organic Antiferromagnet κ-(BEDT-TTF)*$_2$*Cu[N(CN)*$_2$*]Cl*, in *Proceedings of the International Conference on Strongly Correlated Electron Systems (SCES2019)* (Journal of the Physical Society of Japan, Okayama, Japan, 2020).

[28] S. Sumita, M. Naka, and H. Seo, Fulde-Ferrell-Larkin-Ovchinnikov state induced by antiferromagnetic order in κ-type organic conductors, Phys. Rev. Research **5**, 043171 (2023).

[29] L. D. Landau, E. M. Lifšic, L. P. Pitaevskij, and L. D. Landau, *Electrodynamics of Continuous Media*, Second edition revised and enlarged (Pergamon Press, Oxford New York Toronto Sydney Paris Frankfurt, 1984).

[30] S. Bhowal and N. A. Spaldin, Ferroically Ordered Magnetic Octupoles in d-Wave Altermagnets, Phys. Rev. X **14**, 011019 (2024).

[31] R. M. Fernandes, V. S. De Carvalho, T. Birol, and R. G. Pereira, Topological transition from nodal to nodeless Zeeman splitting in altermagnets, Phys. Rev. B **109**, 024404 (2024).

[32] K. Oinuma et al., Spin structure at zero magnetic field and field-induced spin reorientation transitions in a layered organic canted antiferromagnet bordering a superconducting phase, Phys. Rev. B **102**, 035102 (2020).

[33] M. Naka, Y. Motome, and H. Seo, Anomalous Hall effect in antiferromagnetic perovskites, Phys. Rev. B **106**, 195149 (2022).

[34] S. Hayami, Y. Yanagi, and H. Kusunose, Momentum-Dependent Spin Splitting by Collinear Antiferromagnetic Ordering, J. Phys. Soc. Jpn. **88**, 123702 (2019).

[35] T. P. T. Nguyen and K. Yamauchi, *Ab initio* prediction of anomalous Hall effect in antiferromagnetic $CaCrO_3$, Phys. Rev. B **107**, 155126 (2023).

[36] J. M. Williams, A. M. Kini, H. H. Wang, K. D. Carlson, U. Geiser, L. K. Montgomery, G. J. Pyrka, D. M. Watkins, and J. M. Kommers, From semiconductor-semiconductor transition (42 K) to the highest-Tc organic superconductor, κ-$(ET)_2Cu[N(CN)_2]Cl$ (Tc = 12.5 K), Inorg. Chem. **29**, 3272 (1990).

[37] H. Ito, T. Ishiguro, M. Kubota, and G. Saito, Metal-Nonmetal Transition and Superconductivity Localization in the Two-Dimensional Conductor *κ*-$(BEDT\text{-}TTF)_2Cu[N(CN)_2]Cl$ under Pressure, J. Phys. Soc. Jpn. **65**, 2987 (1996).

[38] V. Sunko, C. Liu, M. Vila, I. Na, Y. Tang, V. Kozii, S. M. Griffin, J. E. Moore, and J. Orenstein, *Linear Magneto-Conductivity as a DC Probe of Time-Reversal Symmetry Breaking*, arXiv:2310.15631.

[39] I. Kézsmárki, N. Hanasaki, D. Hashimoto, S. Iguchi, Y. Taguchi, S. Miyasaka, and Y. Tokura, Charge Dynamics Near the Electron-Correlation Induced Metal-Insulator Transition in Pyrochlore-Type Molybdates, Phys. Rev. Lett. **93**, 266401 (2004).

[40] S. Iguchi, S. Kumakura, Y. Onose, S. Bordács, I. Kézsmárki, N. Nagaosa, and Y. Tokura, Optical Probe for Anomalous Hall Resonance in Ferromagnets with Spin Chirality, Phys. Rev. Lett. **103**, 267206 (2009).

[41] R. Shimano, Y. Ikebe, K. S. Takahashi, M. Kawasaki, N. Nagaosa, and Y. Tokura, Terahertz Faraday rotation induced by an anomalous Hall effect in the itinerant ferromagnet $SrRuO_3$, EPL **95**, 17002 (2011).

[42] A. V. Kimel, Th. Rasing, and B. A. Ivanov, Optical read-out and control of antiferromagnetic Néel vector in altermagnets and beyond, Journal of Magnetism and Magnetic Materials **598**, 172039 (2024).

[43] V. V. Eremenko and N. F. Kharchenko, Magneto-optics of antiferromagnets, Physics Reports **155**, 379 (1987).

[44] K. Sato and T. Ishibashi, Fundamentals of Magneto-Optical Spectroscopy, Front. Phys. **10**, 946515 (2022).

[45] D. Faltermeier, J. Barz, M. Dumm, M. Dressel, N. Drichko, B. Petrov, V. Semkin, R. Vlasova, C. Meźière, and P. Batail, Bandwidth-controlled Mott transition in κ-$(BEDT\text{-}TTF)_2Cu[N(CN)_2]Br_xCl_{1-x}$: Optical studies of localized charge excitations, Phys. Rev. B **76**, 165113 (2007).

[46] T. Sasaki, I. Ito, N. Yoneyama, N. Kobayashi, N. Hanasaki, H. Tajima, T. Ito, and Y. Iwasa, Electronic correlation in the infrared optical properties of the quasi-two-dimensional κ-

type BEDT-TTF dimer system, Phys. Rev. B **69**, 064508 (2004).

[47] W. J. Tabor and F. S. Chen, Electromagnetic Propagation through Materials Possessing Both Faraday Rotation and Birefringence: Experiments with Ytterbium Orthoferrite, Journal of Applied Physics **40**, 2760 (1969).

[48] B. Donovan and J. Webster, The Theory of the Faraday Effect in Anisotropic Semiconductors, Proc. Phys. Soc. **79**, 46 (1962).

[49] L. Jastrzębski, Influence of dichroism on Faraday rotation in $YFeO_3$, Phys. Stat. Sol. (a) **21**, 57 (1974).

[50] K. Sato, Measurement of Magneto-Optical Kerr Effect Using Piezo-Birefringent Modulator, Jpn. J. Appl. Phys. **20**, 2403 (1981).

[51] T. Ishibashi et al., Magneto-optical imaging using polarization modulation method, Journal of Applied Physics **100**, 093903 (2006).

[52] S. Iguchi, Y. Ikemoto, H. Kobayashi, H. Kitazawa, H. Itoh, S. Iwai, T. Moriwaki, and T. Sasaki, *Infrared Magneto-Optical Kerr Effect Measurements by Polarization Modulation Method in Anisotropic Magnets*, in *Proceedings of the 29th International Conference on Low Temperature Physics (LT29)*, Vol. 38 (Journal of the Physical Society of Japan, Sapporo, Japan (Hybrid), 2023), p. 011148.

[53] S. Kimura, Infrared spectroscopy under extreme conditions, Physica B: Condensed Matter **329–333**, 1625 (2003).

[54] T. Moriwaki and Y. Ikemoto, BL43IR at SPring-8 redirected, Infrared Physics & Technology **51**, 400 (2008).

[55] S. Kimura and H. Okamura, Infrared and Terahertz Spectroscopy of Strongly Correlated Electron Systems under Extreme Conditions, J. Phys. Soc. Jpn. **82**, 021004 (2013).

[56] F. Kagawa, Y. Kurosaki, K. Miyagawa, and K. Kanoda, Field-induced staggered magnetic moment in the quasi-two-dimensional organic Mott insulator κ-$(BEDT\text{-}TTF)_2Cu[N(CN)_2]Cl$, Phys. Rev. B **78**, 184402 (2008).

[57] H. Seo and M. Naka, Antiferromagnetic State in *κ*-type Molecular Conductors: Spin Splitting and Mott Gap, J. Phys. Soc. Jpn. **90**, 064713 (2021).

[58] A. C. Jacko, E. P. Kenny, and B. J. Powell, Interplay of dipoles and spins in κ-$(BEDT\text{-}TTF)_2X$, where $X$ = $Hg(SCN)_2Cl$, $Hg(SCN)_2Br$, $Cu[N(CN)_2]Cl$, $Cu[N(CN)_2]Br$, and $Ag_2(CN)_3$, Phys. Rev. B **101**, 125110 (2020).

[59] S. M. Winter, K. Riedl, and R. Valentí, Importance of spin-orbit coupling in layered organic salts, Phys. Rev. B **95**, 060404 (2017).

[60] F. J. Kahn, P. S. Pershan, and J. P. Remeika, Ultraviolet Magneto-Optical Properties of Single-Crystal Orthoferrites, Garnets, and Other Ferric Oxide Compounds, Phys. Rev. **186**, 891 (1969).

[61] I. E. Dzialoshinskii, The Problem of Piezomagnetism, Soviet Journal of Experimental and Theoretical Physics **33**, 807 (1957).

[62] T. Moriya, Piezomagnetism in $CoF_2$, Journal of Physics and Chemistry of Solids **11**, 73 (1959).